\documentclass[
amssymb, amsmath,%
aip, jmp,
]{revtex4}

\usepackage{docs}%
\usepackage{bm}%
\usepackage[colorlinks=true,linkcolor=blue]{hyperref}%
\usepackage{amsmath}
\usepackage{amsxtra}
\usepackage{amscd}
\usepackage{amsthm}
\usepackage{amsfonts}
\usepackage{amssymb}
\usepackage{eucal}
\usepackage{mathrsfs}
\usepackage{mathptm}
\usepackage{verbatim}
\usepackage{slashed}

\newcommand{\R}{{\mathbb R}}
\newcommand{\C}{{\mathbb C}}
\newcommand{\Z}{{\mathbb Z}}

\newcommand{\fg}{{\mathfrak g}}

\newcommand{\fl}{{\mathfrak l}}

\newcommand{\0}{{\bar{0}}}
\newcommand{\1}{{\bar{1}}}

\newtheorem{theorem}{Theorem}[section]

\theoremstyle{definition}

\theoremstyle{statement}

\theoremstyle{notation}

\theoremstyle{remark}
\newtheorem{remark}[theorem]{Remark}

\numberwithin{equation}{section}

\expandafter\ifx\csname package@font\endcsname\relax\else
 \expandafter\expandafter
 \expandafter\usepackage
 \expandafter\expandafter
 \expandafter{\csname package@font\endcsname}%
\fi
\begin{document}

\title{Extended Poincar\'e supersymmetry in three dimensions and
supersymmetric anyons}%

\author{M. Chaichian}
\email{Masud.Chaichian@helsinki.fi}
\affiliation
        {Department of Physics, University of Helsinki, P.O. Box 64, 00014 Helsinki, Finland}
 \affiliation{School of Mathematics and Statistics,
        University of Sydney, Sydney, Australia}

\author{A. Tureanu}
\email{Anca.Tureanu@helsinki.fi}
\affiliation
        {Department of Physics, University of Helsinki,
        P.O. Box 64, 00014 Helsinki,Finland}
\affiliation{School of Mathematics and Statistics,
        University of Sydney, Sydney, Australia}

\author{R. B. Zhang}
\email{ruibin.zhang@sydney.edu.au}
\affiliation{School of Mathematics and Statistics,
        University of Sydney, Sydney, Australia}

\begin{abstract}
We classify the unitary representations of the extended
Poincar\'e supergroups in three dimensions. Irreducible
unitary representations of any spin can appear, which
correspond to supersymmetric anyons.
Our results also show that all irreducible unitary
representations necessarily have physical momenta.
This is in sharp contrast to the ordinary  Poincar\'e
group in three dimensions, that admits in addition irreducible
unitary representations with non-physical momenta,
which are discarded on physical grounds.
\end{abstract}


\maketitle


\section{Introduction}\label{sect:introd}

In a series of articles \cite{MT0, MT1, MT2, MT3} in recent years,
Mezincescu and Townsend investigated bosonic strings and superstrings in
target spaces of dimension $3$, which is the only
non-critical dimension where strings can be quantised
consistently. They discovered that the spectra of
$3$-dimensional quantum strings always contain anyons (see
\cite{MT4} for a review).
This showed the relevance of anyons (see, e.g., \cite{FW})
to string theory for the first time.

The stringy anyons are anyons in the $3$-dimensional target space,
which correspond to unitary representations
of the target space Poincar\'e (super)group with spins which are neither integers
nor half integers. Thus, it is essential for the
study of stringy anyons to understand such unitary representations.
In their papers, Mezincescu and Townsend
demonstrated the existence of some unitary representations
for the $N=1$ and $N=2$ Poincar\'e
supergroups in three dimensions. A systematic understanding of the unitary
representations of all the $3$-dimensional Poincar\'e
supergroups is not only useful for the study of supersymmetric anyons, e.g.,
in further developing the Mezincescu-Townsend programme, but is also of interest
in its own right.

Recall that an influential work \cite{S} of Strathdee
from the 1980s classified unitary representations
with integer and half integer spins $\le 2$ of the
extended Poincar\'e supergroups in all dimensions.
His aim was to catalogue the supermultiplets
relevant to supergravity in dimension $4$ and above.
From his point of view, the $3$-dimensional case is
degenerate as there is no graviton.
It appears that the study of unitary representations
of $3$-dimensional extended
Poincar\'e supergroups is still incomplete.

In this note we systematically investigate the
unitary representations of extended Poincar\'e supergroups
in $3$ dimensions. We give a classification
of the irreducible unitary representations
and construct them using the method
of induced representations starting from
unitary representations of the underlying
Poincar\'e group.

The irreducible unitary representations of the
$3$-dimensional Poincar\'e group have been widely studied
(see, e.g., \cite{B}). It is known that unitarity
occurs for both physical and nonphysical momenta.
However, the situation is totally different
for Poincar\'e supergroups: only the positive
energy unitary representations of the
Poincar\'e group with physical momenta can induce
unitary representations of the Poincar\'e
supergroups.

At zero momenta, an irreducible unitary representation of a Poincar\'e supergroup
is nothing more than an irreducible unitary representation of the
underlying Poincar\'e group with the super generators and
central charges all act by zero. In the massless case, all
central charges must act by zero,
and half of the supersymmetries are broken.
We gain a complete understanding of the irreducible unitary
representations for all $N$ in both cases, so do we also
for massive unitary irreducible  representations of the $N=1$
Poincar\'e supergroup. However, in the massive case,
the values of the central charges become crucial in
determining the irreducible unitary representations
of the extended Poincar\'e supergroups. For any given spin,
we will describe the values which allow unitarity.
In particular, for $N=2$ and $3$, the values of the central
charges divided by the mass lie in a unit interval and
a unit ball respectively (see Section \ref{subsect:massive}).

We should point out that a mathematically rigorous
(and technically more demanding) treatment of
induced representations for Lie supergroups was given
in \cite{V}. For general references on the theory of Lie superalgebras,
see \cite{K77} and \cite{Sch}.

\section{Extended Poincar\'e supergroups in three dimensions}\label{sect:super-Poincare}

We choose the metric $\eta=\begin{pmatrix}
-1 & 0 & 0\\
0 & 1 & 0\\
0 & 0 & 1
\end{pmatrix}$
for the $3$-dimensional Minkowski space $M$. The Poincar\'e group is
the semi-direct product $O(1, 2)\rtimes \R^{1, 2}$ acting on $M$ by
\[
(\Lambda, a): \ x^\mu\mapsto \Lambda^\mu_\nu x^\nu + a^\mu,\]
for any $x^\mu=(x^0, x^1, x^2)\in M$. Here
$\Lambda=(\Lambda^\mu_\nu)\in O(1, 2)$ and $a^\mu=(a^0, a^1,
a^2)\in\R^{1, 2}$. The action leaves invariant
$||x-y||^2=(x-y)^\mu(x-y)_\mu$ for all $x, y\in M$.

For physical applications, we are only interested in the connected
component of $O(1, 2)$ containing the identity, defined by
$\{\Lambda\in O(1, 2) \mid \det\Lambda=1, \ \Lambda_0^0>0\}$. This
subgroup can be identified with the quotient group $L=SL(2,
\R)/\Z_2$ of $SL(2, \R)$,  where $\Z_2=\{I, -I\}$. Then $L$ acts on
$x^\mu\in M$ by
\begin{eqnarray}\label{spinor-notation}
\begin{pmatrix}x^0-x^1 & x^2\\ x^2 & x^0+x^1\end{pmatrix}
\mapsto g\begin{pmatrix}x^0-x^1 & x^2\\ x^2 & x^0
+x^1\end{pmatrix} g^T, \quad g\in L.
\end{eqnarray}

Following reference \cite{DM}, we may regard a Lie supergroup as a super
Harish-Chandra pair $(G_\0, \fg)$, where $G_\0$ is an ordinary Lie
group,  $\fg=\fg_\0\oplus\fg_\1$ is a Lie superalgebra which is a
$G_\0$-module with $\fg_\0=Lie(G_\0)$,  such that the $\fg_0$-action
on $\fg$ is the differential of the $G_\0$-action. Thus the extended
Poincar\'e supergroup is the super Harish-Chandra pair with
$G_\0=L\rtimes \R^{1, 2}$ and $\fg$ being the extended super
Poincar\'e algebra in three dimensions.

We largely follow the convention of \cite{S} for the extended
Poincar\'e superalgebra. Take the following $\Gamma$ matrices
\[
\Gamma^0 = i\sigma_2, \quad  \Gamma^1 = \sigma_1 , \quad \Gamma^2 = \sigma_3,
\]
which have real entries.
Then $\{ \Gamma^\mu, \Gamma^\nu\}=2\eta^{\mu \nu}$.
Let $C$ be the charge conjugation matrix, which may be taken as $C=\Gamma^0$.

We denote by $J_{\mu \nu}$ and $P_\mu$ with $\mu, \nu=0, 1, 2$ the
generators of the usual Poincar\'e algebra in three dimensions, where
$J_{\mu \nu}$ are antisymmetric in the indices $\mu$ and $\nu$. The
commutation relations are:
\begin{eqnarray}
\begin{aligned}
{[}J_{\mu \nu}, J_{\sigma \rho}] &= \eta_{\nu \sigma} J_{\mu \rho}
- \eta_{\mu \sigma} J_{\nu \rho}
- \eta_{\nu \rho} J_{\mu \sigma} + \eta_{\mu \rho} J_{\nu \sigma}, \\
[J_{\mu \nu}, P_\sigma] &= \eta_{\nu \sigma} P_\mu -\eta_{\mu \sigma} P_\nu.
\end{aligned}
\end{eqnarray}
In this convention, we have $P_\mu^\dagger = P_\mu$,
$J_{0 \nu}^\dagger = J_{0 \nu}$ and $J_{1 2}^\dagger = -J_{1 2}$.

Let $Q_{i, \alpha}$ be the supercharges, where $\alpha=1, 2$ is the
spinor index, and $i=1, 2, \dots, N$. Denote by $Z_{i j}$ the
central charges. Then
\begin{eqnarray}\label{supercharges}
\{Q_{i, \alpha}, Q_{j, \beta} \} = (\slashed{P} C)_{\alpha \beta}\delta_{i j} +
C_{\alpha \beta} Z_{i j}.
\end{eqnarray}
The central charges are real, i.e., $Z_{i j}^*=Z_{i j}$,  and
satisfy $Z_{i j}= - Z_{j i}$. The supercharges are also real in the
sense that $Q_i = C \overline{Q}_i^T$, where
$\overline{Q}_i=Q_i^\dagger\Gamma^0$. This amounts to $Q_{i,
\alpha}^*=Q_{i, \alpha}$. Set $\Sigma_{\mu
\nu}=\frac{1}{2}[\Gamma_\mu, \Gamma_\nu]$, which are real $2\times
2$-matrices. Then
\begin{eqnarray}
[J_{\mu \nu}, Q_{i \alpha}] &=
\sum_{\beta=1}^2(\Sigma_{\mu \nu})_{\alpha \beta}Q_{i, \beta}.
\end{eqnarray}
Furthermore, $Z_{i j}$,  being central, commute with all the elements in
the Poincar\'e superalgebra; the momentum operators  commute among
themselves and with the supercharges.

It is useful to note that a real $2\times 2$ matrix $\Omega$ belongs
to $SL(2, \R)$ if and only if $C g C^{-1}= g^{-1}$. Furthermore,
\eqref{spinor-notation} can be re-written as
\[
\slashed{x}C \mapsto g \slashed{x}C g^T.
\]

Now under the identification of the $3$-dimensional Lorentz group
with $L$, the Lorentz Lie algebra is identified with the real Lie
algebra $sl(2)$. Denote by $Ad$ the action of $L$ on $\fg$. Then for
any $X\in sl(2)$, we have $Ad_g(X)=g X g^{-1}$,  where $g\in L$. The
action on $P_\mu$ and $Q_{i, \alpha}$ is given by
\begin{eqnarray}\label{actions}
Ad_g (P_\mu)\Gamma^\mu  C = g \slashed{P} C g^T,
\quad Ad_g(Q_{i, \alpha})=(g Q_i)_\alpha,
\end{eqnarray}
where $Q_i$ is the column vector with the two entries $Q_{i, 1}$ and
$Q_{i, 2}$. Obviously, $L$ leaves the central charges invariant.

\section{Induced representations and unitarity}\label{sect:induction}

The method of constructing unitary representations of
the Poincar\'e superalgebra is a generalisation of the well-known
induction method of Wigner and Mackey. A mathematically rigorous
treatment of this generalisation was recently given in reference
\cite{V}.

Note that every irreducible unitary representation of the subgroup
$\R^{1, 2}$ of translations of the Poincar\'e group is of the form
\[
\R^{1, 2} \longrightarrow \C\backslash\{0\}, \quad a\mapsto \exp(i p_\mu a^\mu),
\]
for any fixed three momentum $p_\mu$ in the dual space $\left(\R^{1,
2}\right)^*$ of the subgroup. The differential of the representation
is $P_\mu \mapsto p_\mu$. The operator $P_\mu P^\mu$ in this
representation takes the value $p_\mu p^\mu$, which may be $0, -m^2$
or $m^2$, with $m$ being any fixed real number.

The Lorentz subgroup acts on the subgroup $\R^{1, 2}$ of translations,
and thus also on its dual space $(\R^{1, 2})^*$.
This action changes neither $p_\mu p^\mu$ nor the sign of $p^0$.
Therefore, each orbit of the Lorentz subgroup in $(\R^{1, 2})^*$ is
generated by a momentum,  which is in one of the following standard forms:
\begin{enumerate}
\item  $p_\mu p^\mu=0$:
\[
\begin{aligned}
p^\mu&=0, &\quad  &\text{zero-momentum case},\\
p^\mu&=(\omega, \omega, 0), &\quad  &\text{massless case},\\
p^\mu&=(-\omega, -\omega, 0),
\end{aligned}
\]
where $\omega>0$ is fixed.

\item $p_\mu p^\mu=-m^2$,  with $m>0$:
\[
\begin{aligned}
p^\mu&=(m, 0, 0), &\quad  &\text{massive case},\\
p^\mu&=(-m, 0, 0).
\end{aligned}
\]

\item $p_\mu p^\mu=m^2$,  with $m>0$:
\[
p^\mu=(0, 0, \pm m).
\]
\end{enumerate}

Given a $3$-momentum $p^\mu$ in any of the standard forms listed
above, we let $L_p$ be the subgroup of $L$,  which fixes $p^\mu$. Let
$\fl_p=Lie(L_p)$, which is a subalgebra of $\fg$. Then $\fl_p$, the
momentum operators $P_\mu$, supercharges $Q_{i, \alpha}$ and the
central charges $Z_{i j}$ together span a subalgebra $\fg_0$ of the
Poincar\'e superalgebra $\fg$. The Harish-Chandra pair $(L_p\rtimes
\R^{1, 2}, \fg_0)$ is a subsupergroup of the extended Poincar\'e
supergroup.

Let $\pi_0$ be an irreducible unitary representation of the
subsupergroup $(L_p\rtimes \R^{1, 2}, \fg_0)$ furnished by the complex
Hilbert superspace $V_0(p, z, h)$,  where $z=(z_{i j})$ with $z_{i
j}$ being the eigenvalue of $Z_{i j}$, and $h$ denoting some $L_p$
labels. We construct the Hilbert superspace $V(p, z, h)$ of $V_0(p, z,
h)$-valued $L^2$ functions on $L$ with the following property: for
any $\phi\in V(p, z, h)$,
\[
\phi(gh) = \pi_0(h^{-1}) \phi(g), \quad \forall g\in L, \ h\in L_p.
\]
The action of the extended Poincar\'e supergroup on $V(p, z, h)$ is defined by
\begin{eqnarray}
\begin{aligned}
(h\cdot\phi)(g) &= \phi(h^{-1}g),  \quad h\in L, \\
(a\cdot\phi)(g) &= \exp(i Ad_g(p)_\mu a^\mu)\phi(g), \quad a\in \R^{1, 2}, \\
(Q_{i, \alpha}\cdot\phi)(g) &= \pi_0(Ad_{g^{-1}}(Q_{i, \alpha})) \phi(g), \\
(Z_{i j}\cdot \phi)(g) &= z_{i j} \phi(g), \quad  \forall g\in L.
\end{aligned}
\end{eqnarray}
We shall denote the associated irreducible representation by $\pi$.

Recall that the ordinary Poincar\'e group admits irreducible unitary
representations with non-physical momenta \cite{B}, which are
discarded on physical grounds. This is in sharp contrast to the
present case with supersymmetry, where representations with
non-physical momenta are ruled out by unitarity. More precisely,
\begin{quote}
Poincar\'e supergroups only admit unitary
representations satisfying either $p_\mu p^\mu\le 0$ with $p^0> 0$,
or $p^\mu=0$.
\end{quote}

To prove the claim, we need to show that no unitary representations
exist if $p^\mu$ is one of the standard momenta $(-\omega, -\omega,
0)$, $(-m, 0, 0)$ or $(0, \pm m, 0)$.

If $p^\mu=(-\omega, -\omega, 0)$, we have
\[
\{Q_{i, \alpha}, Q_{j, \beta} \} = -2\omega \delta_{\alpha 2}
\delta_{\beta 2}\delta_{i j}
+C_{\alpha \beta} z_{i j},
\]
when we regard the operators as endomorphisms of $V_0(p, z, h)$. In particular,
\begin{eqnarray}
\{Q_{i, 2}, Q_{j, 2} \} = - 2\omega \delta_{i j}.
\end{eqnarray}
For any nonzero vector $|v\rangle$ in $V_0(p, z, h)$, we have
\[
||Q_{i 2}|v\rangle||^2=\frac{1}{2}\langle v| \{Q_{i, 2}, Q_{i, 2}\}
|v\rangle = - \omega |||v\rangle||^2<0,
\]
contradicting unitarity.

If $p^\mu=(-m, 0, 0)$, we have
\[
\{Q_{i, \alpha}, Q_{j, \beta} \} = -m \delta_{\alpha \beta}\delta_{i j}
+C_{\alpha \beta} z_{i j},
\]
which, in particular,  implies
\begin{eqnarray}\label{negative-energy}
\{Q_{i, \alpha}, Q_{i, \alpha} \}=-m.
\end{eqnarray}

For $p^\mu=(0, \pm m, 0)$, we define the $p$-dependent sign factor
$\epsilon_p=\pm 1$. Then the commutation relations of the
supercharges become
\[
\{Q_{i, \alpha}, Q_{j, \beta} \} = \epsilon_p  (-1)^\alpha m
\delta_{\alpha \beta}\delta_{i j}
+C_{\alpha \beta} z_{i j}.
\]
This, in particular, implies
$
\{Q_{i, 1}, Q_{i, 1} \} = -\{Q_{i, 2}, Q_{i, 2} \}=
-\epsilon_p m.
$
Thus
\begin{eqnarray} \label{tachyon}
\begin{aligned}
\{Q_{i, 1}, Q_{i, 1} \} = - m, &\quad \text{for \ } p^\mu=(0, m, 0), \\
\{Q_{i, 2}, Q_{i, 2} \} = - m, &\quad \text{for \ } p^\mu=(0, -m, 0).
\end{aligned}
\end{eqnarray}

Applying the same arguments used in the $p^\mu=(-\omega, -\omega,
0)$ case to equations \eqref{negative-energy} and \eqref{tachyon},
one can see that no unitary representation exists for the Poincar\'e
supergroup in these cases.

\section{Irreducible unitary representations}\label{sect:unitarity}

Now we consider the irreducible unitary representations of the
Poincar\'e supergroups with a physical momentum $p^\mu$ of
the standard form $p^\mu=0$, $(\omega, \omega, 0)$,  or $(m, 0, 0)$.

\subsection{Zero-momentum case: $p^\mu=0$}\label{subsect:zero}
We have
\[
\{Q_{i, \alpha}, Q_{j, \alpha} \} = 0, \quad \alpha=1, 2.
\]
For any vector $|v\rangle$ in $V_0(p, z, h)$, we have
\begin{eqnarray}\label{norm}
||Q_{i,
\alpha}|v\rangle||^2=\frac{1}{2}\langle v| \{Q_{i, \alpha}, Q_{i, \alpha}\}
|v\rangle=0, \quad \forall i, \alpha.
\end{eqnarray}
Unitarity of $\pi_0$ implies $Q_{i, \alpha}|v\rangle=0$ for all
$|v\rangle$. Hence $\pi_0(Q_{i, \alpha})=0$, for all $i$ and
$\alpha$. This also forces the central charges to be zero.

Now the entire group $L$ leaves $p^\mu$ invariant. Therefore, every
irreducible unitary representation of a Poincar\'e
supergroup is the inflation of an irreducible representation of $L$
by requiring $Q_{i, \alpha}$ and $Z_{i j}$ to act by zero. From the works of Bargmann, Gel'fand and Naimark, and Harish-Chandra
in the 40s and 50s, it is
known \cite{K} that
$SL(2, \R)$ has $3$ series (discrete,
principal and complementary series) of
irreducible unitary representations, beside the trivial representation.
All the irreducible unitary representations are infinite dimensional
except the trivial representation, which is of course $1$-dimensional. Such zero energy-momentum irreducible unitary representations, which have been sometimes called
“spurions” (see the book \cite{Ohnuki}) in imagined analogy with particles, could
perhaps be considered as vacuum states. In this sense, there appears an infinite
degeneracy of vacuum and its non-uniqueness.

\subsection{Massless case: $p^\mu=(\omega, \omega, 0)$}\label{subsect:massless}
The subgroup of $L_p$ consists of the elements $\begin{pmatrix}\pm 1
& 0 \\ c & \pm 1\end{pmatrix}$ in $SL(2, \R)$, which is isomorphic
to $\Z_2\times\R$. Recall that every irreducible unitary
representation of $\R$ is given by $\R\longrightarrow
\C\backslash\{0\}$, $c \mapsto \exp(i t c)$, where $t$ is any fixed
real number. Thus the irreducible unitary representations of $L_p$
are
\[
\begin{aligned}
\phi^{(+; t)}: L_p \longrightarrow \C, &\quad
\begin{pmatrix}\pm 1 & 0 \\ c & \pm 1\end{pmatrix}\mapsto \exp(i t c), \\
\phi^{(-; t)}: L_p \longrightarrow \C, &\quad
\begin{pmatrix}\pm 1 & 0 \\ c & \pm 1\end{pmatrix}\mapsto \pm \exp(i t c).
\end{aligned}
\]

A simple calculation yields
\begin{eqnarray}\label{massless}
\{Q_{i, \alpha}, Q_{j, \beta} \} = 2\omega \delta_{\alpha 2}
\delta_{\beta 2}\delta_{i j}
+C_{\alpha \beta} z_{i j}.
\end{eqnarray}
This, in particular,  implies that
\begin{eqnarray}\label{Clifford}
\{Q_{i, 2}, Q_{j, 2} \} = 2\omega \delta_{i j},
\quad
\{Q_{i, 1}, Q_{i, 1} \} =0.
\end{eqnarray}
A calculation similar to \eqref{norm} shows that
$\pi_0(Q_{i, 1})=0$ for all $i$.
Using this fact in the special case of equation \eqref{massless}
with $\alpha=1$ and $\beta=2$, we obtain $z_{i j}=0$ for all $i, j$.

For any given $h= \begin{pmatrix}\pm 1 & 0 \\ c & \pm 1\end{pmatrix}$ in $L_p$,
we have
\[
\pi_0(Ad_h(Q_{i, 1}))=0, \quad \pi_0(Ad_h(Q_{i, 2}))=\pm
\pi_0(Q_{i, 2}).
\]
Inspecting the first relation in
\eqref{Clifford}, one sees that we may require $V_0(p, z, h)$ to be an irreducible
module for the Clifford algebra spanned by the elements $\frac{Q_{i,
2}}{\omega}$. In fact, every irreducible unitary representation
$\pi_0$ of  $(L_p\rtimes \R^{1, 2}, \fg_0)$ can be obtained by
lifting an irreducible unitary representation of $L_p$ to the
subsupergroup. That is, $\pi_0$ has the following properties:
$\pi_0^{(\pm, t)}(Q_{\alpha, 2})$ is essentially an irreducible
representation of a Clifford algebra as discussed above but
$\pi_0^{(\pm, t)}(Q_{\alpha, 1})=0$, and the restriction of $\pi_0$
to $L_p$ is $\phi^{(\pm; t)}$.

We shall write $\pi^{(\pm; t)}$ for the irreducible unitary
representation $\pi$ of the Poincar\'e supergroup
induced by $\pi_0$, and $V(p, z, t, \pm)$ for $V(p, z, h)$.
For nonzero $t$,  $\pi^{(\pm; t)}$ is the $3$-dimensional
supersymmetric analogue of Wigner's ``infinite spin"
(also termed as ``continuous spin") representations
\cite{BW, W}.

\subsection{Massive case: $p^\mu=(m, 0, 0)$}\label{subsect:massive}

In this case, the subgroup $L_p$, consisting of the elements of
$SL(2, \R)$ satisfying $h h^T=I$, is the special orthogonal group
$SO(2, \R)$ in two dimensions. Its universal cover is
\[
\R\longrightarrow SO(2, \R),\quad \theta\mapsto \begin{pmatrix}
                                                \cos\theta & \sin\theta\\
                                                -\sin\theta &\cos\theta
                                                \end{pmatrix}.
\]
In order to have representations of any ``spin", we need to pass
over to the universal cover and consider its representations
instead. Then every irreducible unitary representation is given by
$\theta\mapsto \exp(i s\theta)$, where the ``spin" $s$ can be any
fixed real number.

We have in this case
\begin{eqnarray}
\{Q_{i, \alpha}, Q_{j, \beta} \} = m \delta_{\alpha \beta}\delta_{i j}
+C_{\alpha \beta} z_{i j}.
\end{eqnarray}
Let $\Psi_i=\frac{Q_{i, 1}+\sqrt{-1} Q_{i, 2}}{\sqrt{2m}}$ and
$\Psi_i^\dagger=\frac{Q_{i, 1}-\sqrt{-1} Q_{i, 2}}{\sqrt{2m}}$; then
the above commutation relations can be re-written as
\begin{eqnarray}\label{Psi}
\{\Psi_i, \Psi_j\} = \{\Psi_i^\dagger, \Psi_j^\dagger\}=0, \quad
\{\Psi_i, \Psi_j^\dagger\}= \delta_{i j} + \sqrt{-1}
\frac{z_{i j}}{m}, \quad \forall i, j.
\end{eqnarray}
For any $h=h_\theta:=\begin{pmatrix}
    \cos\theta & \sin\theta\\
    -\sin\theta &\cos\theta
    \end{pmatrix}$
in $L_p$, we have
\[
Ad_h(\Psi_i)=\exp(-\sqrt{-1}\theta) \Psi_i,
\quad Ad_h(\Psi_i^\dagger)=\exp(\sqrt{-1}\theta) \Psi_i^\dagger.
\]

To analyze the irreducible representations of the subsupergroup
$(L_p\rtimes \R^{1, 2}, \fg_0)$, it is useful to consider
independently the associative superalgebra $\mathcal C$ generated by
the $\Psi_i$ and $\Psi_i^\dagger$ as defined by \eqref{Psi}. The
$\Psi_i$ generate a subsuperalgebra ${\mathcal C}_+$, which has one
irreducible module $\C |0\rangle$ with
\[
\Psi_i |0\rangle=0, \ \forall i.
\]
Construct the induced module
\[
F={\mathcal C}\otimes_{{\mathcal C}_+}|0\rangle,
\]
which has a basis
\[
|{\bf n}\rangle:=\big(\Psi_1^\dagger\big)^{n_1}
\big(\Psi_2^\dagger\big)^{n_2}
\dots\big(\Psi_N^\dagger\big)^{n_N}|0\rangle,
\quad n_i\in\{0, 1\}.
\]

If $N=1$, $F$ is the only unitary module for ${\mathcal C}$.

If $N>1$,  {\em every irreducible $\mathcal C$-module is a quotient}
of $F$. Reducibility / irreducibility of $F$ is controlled by the
real numbers $\frac{z_{i j}}{m}$ related to the central charges, and
unitarity imposes conditions on the possible choices of these
values.

Therefore,  every irreducible module $V_0(p, z, s)$ for $(L_p\rtimes
\R^{1, 2}, \fg_0)$ is a quotient of the module $F(p, z, s)$,
 which is generated by a vacuum vector $|0; p, z, s\rangle$
satisfying
\[
h_\theta|0; p, z, s\rangle= \exp(\sqrt{-1}s\theta)|0; p, z, s\rangle,
\quad \Psi_i |0; p, z, s\rangle=0, \ \forall i,
\]
and the obvious property under the translation subgroup of
the extended Poincar\'e supergroup characterised by the momentum $p$.
A basis for $F(p, z, s)$ is given by
\[
|{\bf n}; p, z, s\rangle:=\big(\Psi_1^\dagger\big)^{n_1}
\big(\Psi_2^\dagger\big)^{n_2}
\dots\big(\Psi_N^\dagger\big)^{n_N}|0; p, z, s\rangle,
\quad n_i\in\{0, 1\}.
\]
Note that $h_\theta |{\bf n}; p, z, s\rangle
=\exp(\sqrt{-1}(s+\sum_in_i)\theta)|{\bf n}; p, z, s\rangle$.

\begin{remark}
What happens here is different from the case of the Clifford algebra,
for which the induced module is always irreducible.
\end{remark}

Let us now consider the different cases in more detail.

We have a complete understanding of the $N=1$ case (see also \cite{MT2}).

For $N=2$, let $w=\frac{z_{12}}{m}$.
The Hermitian form on $V_0$ is normalised so that $\langle
0|0\rangle=1$. Here we have dropped the labels $p, z, s$ from
notations for the sake of simplicity. A straightforward calculation yields $
\langle{1, 1}|{1, 1}\rangle=1 - w^2. $ Therefore, $F$ is irreducible
if and only if $1 - w^2\ne 0$.

In the irreducible case, unitarity requires $w$ to lie in the open
interval $(-1, 1)$ of the real line. This in fact is the necessary
and sufficient condition for $V_0$ to be unitary. This follows from
the fact that both eigenvalues of the matrix
\[
\begin{pmatrix}
\langle{1, 0}|{1, 0}\rangle &\langle{1, 0}|{0, 1}\rangle\\
\langle{0, 1}|{1, 0}\rangle &\langle{0, 1}|{0, 1}\rangle
\end{pmatrix}
=\begin{pmatrix}
 1 & i w\\
 -i w& 1
 \end{pmatrix}
\]
are positive under the condition on $w$.

If $w=\pm 1$, $F$ is reducible with a $2$-dimensional maximal
submodule spanned by $(\Psi_1^\dagger+ iw \Psi_2^\dagger)|0\rangle$
and $\Psi_1\Psi_2^\dagger|0\rangle$. The irreducible quotient $V_0$
is isomorphic to the Fock space of the fermionic operators
\[
\Psi=\frac{1}{2}\left(\Psi_1+iw \Psi_2\right), \quad \Psi^\dagger
=\frac{1}{2}\left(\Psi_1^\dagger-iw \Psi_2^\dagger\right),
\]
and thus is automatically unitary. In this case,
half of the supersymmetry is broken.

Therefore, the massive irreducible unitary representations of the
$N=2$ extended Poincar\'e supergroup with fixed $s$ are parametrized by the
closed interval $[-1, 1]$. The irreducible unitary representations
$V(p, z, s)$ with $w$ lying on the boundary arose in \cite{MT1}, and
the supermultiplets correspond to semions when $s=-\frac{1}{4}$ or
$s=-\frac{3}{4}$.

In the $N=3$ case, the irreducible massive unitary representations
with fixed $s$ are parametrized by the unit ball $B:=\{(w_1, w_2,
w_3)\in\R^3\mid w_1^2 +w_2^2 + w_3^2\le 1\}$, where
$w_1=\frac{z_{12}}{m}$, $w_2=\frac{z_{13}}{m}$ and
$w_3=\frac{z_{23}}{m}$. Furthermore, $F$ is reducible if and only if
$w$ lies on the boundary $\partial B$, which is the sphere $ w_1^2
+w_2^2 + w_3^2=1$. In this case, $\dim_\C V_0=4$ with half of the
dimensions even and half odd, independent of the actual position of
$w$ in the sphere. One can verify these results in much the same way
as in the $N=2$ case and  continue the study for the extended Poincar\'e
supergroups  with  higher values of $N$.

\section{Conclusion}
We have classified the irreducible unitary representations
of the extended Poincar\'e supergroups in $3$ dimensions.
Since unitary representations are always completely reducible,
this achieves a classification of all the unitary representations.
Irreducible unitary representations with arbitrary spins appear,
which correspond to supersymmetric anyons.
We have also shown that all the irreducible unitary representations
have physical energy-momenta, an observation which remains valid in other
dimensions too, as can be shown by similar arguments as those used here.
This is essentially different from the case of the ordinary Poincar\'e group,
which admits unitary representations also with nonphysical momenta \cite{B}.

\bigskip
\noindent{\bf Acknowledgements}:
We are grateful to Paul Townsend for many useful discussions and for encouragement.
Financial support from the Academy of Finland (Projects No.136539 and No.140886)
and the Australian Research Council (Projects DP0772870 and DP0986551)
is acknowledged.

\end{document}